\documentclass[%
noreprint,
 amsmath,amssymb,
aps,
]{revtex4-2}

\usepackage{natbib}

\usepackage{graphicx}% Include figure files
\usepackage{dcolumn}% Align table columns on decimal point
\usepackage{bm}% bold math
\usepackage{hyperref}% add hypertext capabilities
\usepackage{upgreek}
\newcommand\mi{\mathrm{i}}
\newcommand\me{\mathrm{e}}
\newcommand\const{\text{const}}
\newcommand\pp{\uppi}
\newcommand{\dif}{\mathrm{d}}
\DeclareMathOperator{\diag}{diag}
\usepackage{braket}
\usepackage{color,soul}

\begin{document}

\title{Entropy and Topology of Regular Black Holes}

\author{Chen Lan}
 \email{lanchen@nankai.edu.cn}
\author{Yan-Gang Miao}
 \email{miaoyg@nankai.edu.cn}
 \altaffiliation{corresponding author.}

   \affiliation{
   School of Physics, Nankai University, 94 Weijin Road, Tianjin 300071, China.
   }

\date{\today}

\begin{abstract}
We calculate the entropy of spherically symmetric regular black holes
by the path integral
and Noether-charge method.
Both methods provide an evidence that the entropy of regular black holes
should be proportional to quarter of area,
and there is no violation of entropy/area law at all.
\end{abstract}

\maketitle

%\tableofcontents

\section{\label{sec:intr}Introduction}

Because it is simpler than the path integral or Noether-charge approach \cite{Gibbons:1976ue,Wald:1993nt},
the computing method of entropy 
\begin{equation}
S_c=\int_{r_{\rm ext}}^{r_{\rm H}} (\dif M \beta)_c,\qquad
\beta=1/T
\end{equation}
is widely used for investigating the thermodynamics in regular black holes (RHBs), e.g.\ \cite{Myung:2007qt,Myung:2008kp,Spallucci:2008ez,Miao:2015npc,Nam:2018sii,Lan:2020wpv,Kumara:2020ucr}.
This method is based on the first law of thermodynamics and
the direct correspondence between internal energy and BH mass, $\braket{U}=M$.
%Nevertheless, if the internal energy derived from the classical action $I_{\rm cl}$ 
%\begin{equation}
%\braket{U}=\frac{\partial I_{\rm cl}}{\partial \beta}
%\end{equation}
%does not equal to $M$, 
%it will lead to several problems.
As we pointed in previous work \cite{Lan:2021ngq},
the conditional entropy $S_c$ for RBHs is incorrect
because it is not consistent with the result computed from the first principle.
In this work, we discuss the entropy for more types of BHs  
by two different ways mentioned above, i.e.\ the path integral and Wald's method.

Let us start with RHBs with a single shape function
\begin{equation}
\label{eq:original}
f(r)=1-\frac{2M}{r}\sigma\left(r, M, \alpha_i\right),
\end{equation}
At first, 
we rewrite $\sigma$ and its derivatives via the curvature invariants 
\begin{align}
\label{eq:sigma}
\frac{\sigma(r)}{r^3}&=\frac{1}{24M}\left(R+3 \sqrt{2E}+2 \sqrt{3W}\right),\\
\frac{\sigma'(r)}{r^2}&=\frac{1}{8M}\left(R+\sqrt{2E}\right),\\
\frac{\sigma''(r)}{r}&=\frac{1}{4M}\left( R-\sqrt{2E}\right).
\end{align}
The regularity at $r=0$ implies
$\lim_{r\to0} \sigma/r^3=\const$
from the perspective of finite curvatures. 
For the case, when $\sigma$ has a power expansion around zero,
$\sigma=\sum_{n=3}^N a_n r^n$, where $N$ could be infinite,
we find the extreme case
\begin{equation}
R\sim O(1),\qquad
E\sim O(r^2),\qquad
W\sim O(r^2)
\end{equation}
and  $E\sim W\sim R^2$. 
For generalized case,
the above asymptotic behaviors can be recast in the following
\begin{equation}
\lim_{r\to 0}R=\const,\qquad
\lim_{r\to 0}\frac{E}{r^2}=\const,\qquad
\lim_{r\to 0}\frac{W}{r^2}=\const.
\end{equation}
In other words, the leading terms of $\sigma$  and its derivatives should be
\begin{equation}
\label{eq:asymptotic-sigma}
 \sigma \sim \frac{R^{(0)}}{24M}r^3,
 \qquad
\sigma'\sim\frac{R^{(0)}}{8M} r^2,
\qquad
\sigma''\sim\frac{R^{(0)}}{4M} r,
\end{equation}
as $r\to 0$, 
where $R^{(0)}$ is the limit of Ricci curvature at zero.
As a result, the metric of RHBs tends to flat as the radius approaches to zero,
which, in turn, provides the same situation of geodesic completeness as the zero point of flat spacetime.
In other words, even though the null or timelike geodesic arrives at $r=0$ in finite affine parameter, 
it does not affect the completeness of the spacetime, 
because the zero point is not even singular point of metric at all.

We end this introductory section by emphasizing an advantage of new representation
obtained by replacing $\sigma$ in Eq.~\eqref{eq:original} by Eq.~\eqref{eq:sigma},
\begin{equation}
\label{eq:shape-new}
f(r)=1-\frac{\lambda}{3}r^2, \qquad
4\lambda=R+3  \sqrt{2E}+2  \sqrt{3W}.
\end{equation}
This representation provides a simple way to check whether a BH model is regular
by directly observing the asymptotic behavior of $\lambda r^2$ around zero.
In particular, when $\lambda r^2$ is a rational function of $r$,
the balance of the lowest power of $r$ aprearing in numerator and 
denominator for a RHB should not be greater than two.

\section{\label{sec:action} The classical action and partition functions}

To observe the entropy, we start with the partition function of gravitational sector
in the framework of Einstein's gravity.
After Wick rotation $t\to -\mi\tau$, 
it can be derived at zero-loop approximation \cite{Gibbons:1976ue}
\begin{equation}
\mathcal{Z}=\int \mathcal{D} g\; \me^{-I}\approx  \me^{-I_{\rm cl}},
\end{equation}
where the Euclidean action $I_{\rm cl}$ consists of three parts \cite{Poisson:2009pwt}
\begin{equation}
\label{eq:class-action}
I_{\rm cl} = I_{\rm EH}+I_{\rm GHY}-I_0+\dots.
\end{equation}
The Einstein-Hilbert action $I_{\rm EH}$, 
Gibbons-Hawking-York boundary term $I_{\rm GHY}$
\begin{equation}
\label{eq:action}
I_{\rm EH} =-\frac{1}{16\pp}\int_{\mathcal{M}}\dif^{4}x\sqrt{-g}\,R,
\qquad
I_{\rm GHY} =-\frac{1}{8\pp} \int_{\partial\mathcal{M}}\dif^{3}x\sqrt{-h}\, K,
\end{equation}
and the counterterm $I_0$ for regularizing the IR divergence appearing in $I_{\rm GHY}$
\begin{equation}
\label{eq:counterterm}
I_0 =-\frac{1}{8\pp} \int_{\partial\mathcal{M}}\dif^{3}x\sqrt{-h}\, K_0.
\end{equation}
where $R$ is Ricci curvature of bulk, $K$ and $K_0$ are extrinsic curvature of surface
and background reference separately. 
The ellipsis omits the matter sectors, 
which will not be involved in calculating of internal energy and entropy of RBHs.
This is related to the so-called Christodoulou-Ruffini irreducible mass \cite{Christodoulou:1970wf,Christodoulou:1972kt},
and we will call the action without matter sector as {\em irreducible} part.
The detailed explanation will be given in the App.~\ref{app:rnbh} based on Reissner-Nordstr\"om BH.

\subsection{Regular black holes with single shape function}

For the spherically symmetric RBHs 
\begin{equation}
g_{\mu\nu}= \diag\!\left\{
-f(r), 
f^{-1}(r),
r^2,
r^2\sin^2(\theta)
\right\}
\label{eq:metric}
\end{equation}
 with single shape function Eq.~\eqref{eq:original},
one can compute the corresponding curvature invariants
\begin{equation}
\label{eq:curvature}
R=\frac{2 M}{r^2}  \left(2 \sigma '+r \sigma ''\right),\qquad
K=\frac{2 r  -M r \sigma '-3 M \sigma}{r^{3/2} \sqrt{r-2 M \sigma }}
\end{equation}
and extrinsic curvature $K_0=2/r$ of the surface embedded in flat spacetime.
Here the prime denotes the derivative with respect to $r$.
Since $\sigma (r)\sim O(r^n)$ with $n\ge 3$ as $r\to 0$,
the derivatives of $\sigma$ can not globally vanish.
Thus the EH part will make a contribution,
which is different from that one of traditional singular models, such as Schwarzschild BH.
Considering the measure of bulk $\sqrt{-g} = r^2 \sin (\theta )$
and surface
$\sqrt{-h}= r^{3/2} \sin (\theta ) \sqrt{r-2 M \sigma }$,
the integrals Eq.~\eqref{eq:action}  can be taken off
\begin{equation}
I_{\rm EH}= -\frac{ M }{2}\beta\left(r \sigma'+\sigma\right)\Big |_{r_{\rm H}}^{\infty},
\end{equation}
where the integration of Euclidean time $\tau$ runs from $0$ to $\beta=1/T$
and radial $r$ from outer horizon $r_{\rm H}$ to $\infty$.
On the other side,
taking into account that the RHBs are asymptotic to Schwarzschild BH at infinity \cite{Lan:2021ngq}
\begin{equation}
\lim_{r\to\infty}\sigma =1,\qquad
\lim_{r\to\infty} r\sigma' =0,
\end{equation}
we obtain
\begin{equation}
I_{\rm EH}= 
\frac{\beta}{2}   \left(r_{\rm H}-M\right)-\pp  r_{\rm H}^2,
\end{equation}
and
the GHY term is
\begin{equation}
I_{\rm GHY} =
\frac{M }{2}\beta\left(r \sigma'+3 \sigma\right)
-r \beta\sim
\frac{3M }{2}\beta
-r \beta,
\end{equation}
which is divergent as a manner of $r$ as $r\to\infty$ 
and regularized by the couterterm
\begin{equation}
I_0 =-r\beta\sqrt{1-\frac{2 M \sigma}{r}}
\sim -r\beta+M\beta+O\left(r^{-1}\right).
\end{equation}
Summarizing all above contributions,
we arrive at classical Euclidean action

%\begin{equation}
%I_{\rm cl}= \frac{r_{\rm H}+2M r_{\rm H} \sigma'_{\rm H}}{4T}
%\end{equation}

%\begin{equation}
%F=-T \ln(Z)=\frac{1}{4}\left(r_{\rm H}+2M r_{\rm H} \sigma'_{\rm H}\right)
%\end{equation}

%\begin{equation}
%T=\frac{1}{4 \pp  r_{\rm H}}-\frac{M \sigma '_{\rm H}}{2 \pp  r_{\rm H}}
%\end{equation}

%\begin{equation}
%S=-\left(\frac{\partial_M F}{\partial_M T}\right)_{r_{\rm H}}=\pp r_{\rm H}^2
%\end{equation}

\begin{equation}
\label{eq:class-action}
I_{\rm cl}=-\pp r_{\rm H}^2+\frac{r_{\rm H}}{2}
\beta.
\end{equation}
Then the internal energy and entropy of RBHs are clearly in order 
\begin{equation}
\label{eq:area-law}
\braket{U}=\frac{\partial I_{\rm cl}}{\partial \beta}=\frac{r_{\rm H}}{2}
,\qquad
S=\beta\braket{U}-I_{\rm cl}
=\pp r_{\rm H}^2,
\end{equation}
where we have supposed that 
$\beta$ and $r_{\rm H}$ are two independent variables, 
which is equivalent to 
a \emph{isometric}  process $V_{\rm th}\equiv 4 \pp r_{\rm H}^3/3=\const$.
Meanwhile, it is worth to note that the internal energy $\braket{U}$ given here 
refers to ``the matter-independent energy''
equivalent to the irreducible mass \cite{Christodoulou:1970wf}.

Having the classical action Eq.~\eqref{eq:class-action} at hand,
one can also calculate the thermal pressure $P_{\rm q}$ by
\begin{equation}
P_{\rm q}=-\frac{\partial \left(\beta^{-1} I_{\rm cl}\right)}{\partial V_{\rm th}}
=-\frac{\beta -4 \pp  r_{\rm H}}{8 \pp  \beta  r_{\rm H}^2},
\end{equation}
and it is reasonable to ask wether this pressure consists with 
the pressure $P_{\rm G}\equiv G^r_{\; r}/(8\pp)$ generated by Gliner vacuum (or vacuum-like medium) at horizon, 
which is discussed in \cite{Lan:2021ngq}.
After calculating the Einstein tensor of metric Eq.~\eqref{eq:metric},
one can read off the pressure of Gliner vacuum from $r$-$r$ component,
\begin{equation}
P_{\rm G}=-\frac{M \sigma '\left(r_{\rm H}\right)}{4 \pp  r_{\rm H}^2},
\end{equation}
while $\sigma '\left(r_{\rm H}\right)$ can be represented via $\beta$, i.e.
\begin{equation}
\sigma '\left(r_{\rm H}\right)= \frac{\beta -4 \pp  r_{\rm H}}{2 \beta  M}.
\end{equation}
Therefore one can verify that the Gliner pressure is exactly the thermal pressure $P_{\rm q}$.
Not only for RHBs, 
but also for all traditional singular BHs, such as RN-AdS BH,
a thermal pressure can be derived from classical action,
if the thermal volume $V_{\rm th}$ is regarded as BH's volume,
and this thermal pressure must consist with Gliner pressure.

\subsection{The isometric process}

To derive the entropy/area law Eq.~\eqref{eq:area-law},
we have supposed that $r_{\rm H}$ is independent with $\beta$.
To understand its meaning, 
let us consider a typical example,
Hayward BH
with shape function \cite{Hayward:2005gi}
\begin{equation}
\label{eq:hayward}
f_e = 1-\frac{2M}{r}\frac{ r^3}{r^3+e^3}
\quad
\text{or}\quad
f_\Lambda=1-\frac{2M}{r}\frac{r^3}{r^3+\frac{6 M}{\Lambda }}.
\end{equation}
These two parameterizations are mathematically equivalent, 
but have different physical explanations. 
We are primarily going to use the former one in the following.

 Hayward BH is of  two ``constraints'', 
one comes from $f(r_{\rm H})=0$, 
the other is $\beta=4\pp/f'(r_{\rm H})$,
\begin{equation}
\label{eq:constraint}
1-\frac{2 M r_{\rm H}^2}{e^3+r_{\rm H}^3}=0,\qquad
\beta = \frac{2 \pp  \left(e^3+r_{\rm H}^3\right)^2}{M r_{\rm H} \left(r_{\rm H}^3-2 e^3\right)}.
\end{equation}
The ``constraints'' imply that the four variables $\beta$, $M$, $e$ and $r_{\rm H}$
are related to each other.
In other words, there are only two of them are independent in this system.
Since the shape function depends on two parameters $M$ and $e$,
this character is obvious. 
Now, instead of choosing original $M$ and $e$ as independent variables, 
we prefer to take two others, $\beta$ and $r_{\rm H}$.
This is what we mean by ``isometric process" in previous subsection.

The corresponding sectors are separately
\begin{equation}
\begin{split}
I_{\rm EH}=-\frac{\beta  e^3 M \left(e^3-2 r_{\rm H}^3\right)}{2 \left(e^3+r_{\rm H}^3\right)^2}&,\qquad
I_{\rm GHY}=-\frac{\beta  r \left[2 e^3 r^2 (2 r-3 M)+2 e^6+r^5 (2 r-3 M)\right]}{2 \left(e^3+r^3\right)^2},\\
&I_0=-\beta  r \sqrt{1-\frac{2 M r^2}{e^3+r^3}}.
\end{split}
\end{equation}
Thus the total contribution is
\begin{equation}
I_{\rm cl}=\frac{\beta  M \left(4 e^3 r_{\rm H}^3+r_{\rm H}^6\right)}{2 \left(e^3+r_{\rm H}^3\right)^2}.
\end{equation}
To reduce the parameters,
we apply the constraints Eq.~\eqref{eq:constraint} to replace $M$ and $e$
\begin{equation}
M\to \frac{3 \beta  r_{\rm H}}{4 \left(\beta +2 \pp r_{\rm H}\right)},\qquad
e^3\to -\frac{r_{\rm H}^3 \left(4 \pp  r_{\rm H}-\beta \right)}{2 \left(\beta +2 \pp  r_{\rm H}\right)}
\end{equation}
and obtain a reexpressed action
\begin{equation}
\label{eq:action-hayward}
I_{\rm cl}=\frac{\beta  r_{\rm H}}{2}-\pp  r_{\rm H}^2.
\end{equation}
Since $\beta$ and $r_{\rm H}$ are independent, we can conclude that
\begin{equation}
\label{eq:hayward-result}
\braket{U}_{\rm iso}\equiv \frac{\partial I_{\rm cl}}{\partial \beta}=\frac{r_{\rm H}}{2},\quad
S_{\rm iso}\equiv \beta\braket{U}-I_{\rm H}=\pp r_{\rm H}^2.
\end{equation}

Furthermore, the choice of independent variables corresponds to the different processes,
and may lead to different results.
To see it clearly, let us now take $\beta$ and $M$ as independent variables, 
which is equivalent to keep $M$ being constant.
Now the internal energy should be calculated by
\begin{equation}
\braket{U}_M = \frac{\partial I_{\rm cl}}{\partial \beta}+
\frac{\partial I_{\rm cl}}{\partial r_{\rm H}} \left( \frac{\partial r_{\rm H}}{\partial \beta}\right)_M.
\end{equation}
The results show the difference from Eq.~\eqref{eq:hayward-result}
%The classical action becomes
%\begin{equation}
%I_{\rm cl}=\frac{2 \beta ^2 M (3 \beta -16 \pp  M)}{(8 \pp  M-3 \beta )^2}.
%\end{equation}
%The internal energy and entropy are separately in terms of $r_{\rm H}$ and $e$
%\begin{equation}
%\braket{U}=-\frac{2 \beta  M \left(9 \beta ^2+256 \pp ^2 M^2-72 \pp  \beta  M\right)}{(8 \pp  M-3 \beta )^3},\qquad
%S=-\frac{256 \pp ^2 \beta ^2 M^3}{(8 \pp  M-3 \beta )^3}
%\end{equation}
%or in terms of $r_{\rm H}$ and $e$
\begin{equation}
\braket{U}_M=\frac{r_{\rm H}}{2}+\frac{3 e^3 r_{\rm H} \left(2 e^3-r_{\rm H}^3\right)}{4 \left(e^3+r_{\rm H}^3\right)^2},\qquad
S_M=\pp  r_{\rm H}^2-\frac{3 \pp  e^3 r_{\rm H}^2}{e^3+r_{\rm H}^3}.
\end{equation}
Alternatively, if the charge $e$ is supposed to be fixed,
one has
\begin{equation}
\braket{U}_e=\frac{r_{\rm H}}{2}+
\frac{3 e^3 r_{\rm H} \left(2 e^3-r_{\rm H}^3\right)}{20 e^3 r_{\rm H}^3+4 e^6-2 r_{\rm H}^6}
,\qquad
S_e=\pp  r_{\rm H}^2
-\frac{6 \pp  e^3 r_{\rm H}^2 \left(e^3+r_{\rm H}^3\right)}{10 e^3 r_{\rm H}^3+2 e^6-r_{\rm H}^6}.
\end{equation}
On the other side, to highlight the difference, 
let us write down
the conditional entropy
\begin{equation}
S_c = 
\int^{r_{\rm H}}_{r_{\rm ext}} 
\dif \tilde r_{\rm H}
\beta
\left(
\frac{\partial M}{\partial \tilde r_{\rm H}}
\right)_{e}
=\pp  r_{\rm H}^2-\frac{2 \pp  e^3}{r_{\rm H}}.
\end{equation}
It notes that the conditional entropy is not consistent with any of above three cases.
%The root of the problem is equalling the internal energy $\braket{U}$
%and mass parameter $M$, 
%which actually are not equivalent.

\subsection{The isometric entropy of Gauss-Bonnet theory}

To see how the isometric process affects the results of {\em singular} BHs,
let us see one more example,  
the well-known Gauss-Bonnet (GB) BH.
We are going to show its entropy in various processes.
The Euclidean action of Einstein-Gauss-Bonnet (EGB) sector is
\begin{equation}
\label{eq:action-egb}
I_{\rm EGB} =-\frac{1}{16\pp}\int\dif^{d} x \sqrt{-g}(R+
\alpha B),\qquad
B=K-4R_2 +R^2,
\end{equation}
where $\alpha$ is coupling constant, while $B$ is GB scalar.
Moreover there is an additional boundary term due to GB correction
besides GHY term
\cite{Myers:1987yn,Myers:1988ze,Brihaye:2008xu},
\begin{equation}
I_{M}=-\frac{\alpha}{16\pp}
\int_{\partial\mathcal{M}} \dif^{d-1} x
\sqrt{-h}\, \mathcal{C},
\end{equation}
where $\mathcal{C}=J-2G_{a b} K^{a b}$, the definition of $J$ can be found in Ref.~\cite{Brihaye:2008xu}.
Then applying the equation of motion
\begin{equation}
R_{\mu\nu}-\frac{1}{2}g_{\mu\nu} R +\alpha \mathcal{H}_{\mu\nu}=0,
\end{equation}
one can rewrite the EGB sector by
\begin{equation}
\label{eq:onshell-egb}
I_{\rm EGB} =\frac{1}{8(d-4)\pp}
\int\dif^{d} x 
\sqrt{-g}\,
R.
\end{equation}
%where we it is easy to find the trace of GB tensor $\mathcal{H}=(4-d)B/2$, 
%thus $R=-\alpha(d-4)B/(d-2)$.
The shape function of BH solution can be cast in the following \cite{Cai:2001dz}
\begin{equation}
f=1+\frac{r^2}{2\tilde{\alpha }}
\left(1-\sqrt{
1+\frac{64\pp\tilde{\alpha }M }{(d-2)A_{d-2} r^{d-1}}
}\right)
\end{equation}
with redefined coupling constant $\tilde\alpha=\alpha(d-3)(d-4)$.
%If one concentrates on high dimensional case $d>4$,
%it is not difficult the find the asymptotic behaviors of $f$ and its direvative
%\begin{equation}
%\label{eq:horizon}
%M= \frac{(d-2) A_{d-2} }{16 \pp } r_{\rm H}^{d-5} \left(\tilde\alpha+r_{\rm H}^2\right)
%\end{equation}
%\begin{equation}
%\label{eq:horizon2}
%\tilde\alpha =- r_{\rm H}^2+
% \frac{16 \pp  M r_{\rm H}^{5-d}}{ (d-2) A_{d-2}}
%\end{equation}
%\begin{equation}
%\label{eq:temp-gb}
%\beta^{-1}= \frac{\tilde{\alpha } (d-5)+(d-3) r_{\rm H}^2}{4 \pp r_{\rm H} \left(2 \tilde{\alpha }+r_{\rm H}^2\right)}
%\end{equation}
%\begin{equation}
%1-f\sim \frac{16\pp M}{(d-2)A_{d-2} r^{d-3}}
%+O\left(1/r^{d-2}\right),\qquad
%f'\sim \frac{(d-3)16\pp M}{(d-2)A_{d-2} r^{d-2}}
%+O\left(1/r^{d-1}\right)
%\end{equation}
The corresponding contributions are separately 
\begin{equation}
I_{\rm EGB} =
\frac{2M\beta}{(d-4) (d-2)}
-
\frac{A_{d-2}\beta}{8(d-4)\pp }
\left[
(d-2) r_{\rm H}^{d-3}
-r_{\rm H}^{d-2} f'(r_{\rm H})
\right],
\end{equation}
\begin{equation}
I_{\rm GHY}=-\frac{(d-2)A_{d-2}\beta}{8\pp }
r_\infty^{d-3} 
+\frac{d-1}{d-2}M\beta,
\end{equation}
where $r_\infty$ denotes the large value of radius, 
and signifies IR divergence.
The contribution of GB term is
\begin{equation}
I_{\rm M}=-\frac{(d-2)\tilde{\alpha}A_{d-2}\beta}{24\pp} 
 r_\infty^{d-5},    
\end{equation}
which is subtracted completely by the counterpart with flat $\mathcal{C}_0=2(d-4) (d-3) (d-2)/(3 r^3)$, 
i.e.\ the GB term does not contribute to the final results.
The IR divergence in GHY term is subtracted by a similar term with
$K_0=(d-2)/r$,
\begin{equation}
I_0=-\frac{(d-2) \beta A _{d-2}r^{d-3}_\infty}{8 \pp  }+
M\beta.
\end{equation}
Summarizing all above formulas, we arrive at
\begin{equation}
I_{\rm lc}=I_{\rm EGB}+I_{\rm GHY}-I_0
%=
%\frac{A_{d-2} r_{\rm H}^{d-3} \left(2-d+4 \pp  T r_{\rm H}\right)}{8 \pp  (d-4) T}+\frac{M}{(d-4) T}
=\frac{A_{d-2} r_{\rm H}^{d-5} }{16 \pp  (d-4)}
\left[(d-2) \left(\tilde\alpha-r_{\rm H}^2\right)\beta+8 \pp r_{\rm H}^3\right],
\end{equation}
which is exactly the same classical action obtained in the seminal work of Myers and Simon \cite{Myers:1988ze}. 
The internal energy then can be calculated by fixing the coupling constant 
\begin{equation}
\braket{U}_\alpha=\frac{\partial I_{\rm lc}}{\partial \beta}+
\frac{\partial I_{\rm lc}}{\partial r_{\rm H}}
\left(\frac{\partial r_{\rm H}}{\partial \beta}\right)_{\alpha},
\end{equation}
which gives a direct correspondence between internal energy and BH mass $\braket{U}_\alpha=M$.
This is the reason that why the conditional entropy $S_c$ provides the same result 
under the fixed coupling constant $\alpha$ \cite{Cai:2001dz}.
Meanwhile the entropy is calculated by $S_{\alpha}=\beta M - I_{\rm cl}$, 
\begin{equation}
\label{eq:egb-entropy}
S_\alpha=\frac{A_{d-2} r_{\rm H}^{d-2} }{4 }
\left[1
+\frac{ (d-2)}{(d-4) }
\frac{2 \tilde \alpha}{r_{\rm H}^2}
\right].
\end{equation}
The deviation is shown in second term of square bracket.
Although it breaks the entropy/area law,
it should have the same name with the one in Schwarzschild BH, 
i.e.\ Bekenstein-Hawking entropy,
because it is derived by same method originated in Ref.~\cite{Gibbons:1976ue}

Choosing $\alpha$ being constant is completely reasonable,
but let us see what the isometric process gives for this singular BH
\begin{equation}
\braket{U}_{\rm iso}=
\frac{A_{d-2} (d-2) r_{\rm H}^{d-7} 
}{8 \pp  (d-4) (d-1)}
\left[(d-5) \tilde{\alpha }^2+2 (d-3) \tilde{\alpha } r_{\rm H}^2-r_{\rm H}^4\right],
\end{equation}
\begin{equation}
S_{\rm iso}=
\frac{A_{d-2} r_{\rm H}^{d-6} 
}{4 (d-4) (d-1)}\left[4 (d-2) \tilde{\alpha }^2+4 (d-2) \tilde{\alpha } r_{\rm H}^2-d r_{\rm H}^4\right].
\end{equation}
The entropy/area law  is violated as well.

\section{Conformal regular black holes}

Now we turn to consider the RHBs in the framework of conformal invariant model,
whose Euclidean action is \cite{Dabrowski:2008kx}
\begin{equation}
\label{eq:action-conformal}
I_{\rm conf}=-\frac{1}{2}\int \dif^4 x \sqrt{-g} \phi \left(
\frac{1}{6}R \phi -\Box \phi
\right).
\end{equation}
Since the equation of motion is $R = 6 \Box \phi/\phi$,
this sector does not contribute to classical action in general. 
The contributions from boundary term and background reference are
\cite{Dyer:2008hb}
\begin{equation}
I_{\rm GHY}=-\int \dif^3 x \sqrt{-h} 
\frac{\phi^2 }{6}K,\qquad
 I_{0}=-\int \dif^3 x \sqrt{-h} 
\frac{\phi^2 }{6}K_0. 
\end{equation}
Since under the transformation $\tilde h_{ij} =\Omega^2 h_{ij}$, 
one has \cite{Kiefer:2017nmo}
\begin{equation}
\sqrt{-\tilde h} =\Omega^3\sqrt{- h},\qquad
\tilde K = \frac{1}{\Omega} \left(
K + 3\mathcal{L}_{\bm n} \Omega
\right),
\end{equation}
where $\mathcal{L}_{\bm n} \Omega\equiv n^\mu \partial_\mu \Omega$ 
is Lie derivative of $\Omega$ along the normal vector field $\bm{n}$.
Thus the actions with tildes become
\begin{align}
\tilde I_{\rm GHY} &=-\int \dif^3 x \sqrt{-\tilde h} 
\frac{\tilde \phi^2 }{6}\left[\tilde K 
-3 n^\mu\partial_\mu \ln(\Omega)
\right],\\
\tilde I_{0}&=-\int \dif^3 x \sqrt{-\tilde h} 
\frac{\tilde\phi^2 }{6}\left[\tilde K_0
-3 n_0^\mu\partial_\mu \ln(\Omega)
\right].
\end{align}
It implies that both $I_{\rm GHY}$ and $I_{0}$ are not conformal invariant independently,
but their difference vanishes
\begin{equation}
\Delta I =\frac{1}{2} \int \dif^3 x \sqrt{-\tilde h} \, \tilde\phi^2 (n^{\mu}-n^{\mu}_0)\partial_\mu \ln(\Omega)
\end{equation}
as $r\to \infty$, because $n^{\mu}$ is asymptotic to $n^{\mu}_0$.
Thus one can make a conclusion that a regular black hole,
received by a conformal change to another singular BH,
should have the same classical action with that singular one.

Now let us study a specific regular model.
 The metric reads $\tilde g_{\mu\nu}=\Omega^2 g_{\mu\nu}$, 
 where $g_{\mu\nu}$ is metric of Schwazschild black hole, 
 and the scale factor is $\Omega= (1+L^2/r^2)^n$ \cite{Toshmatov:2017kmw}.
The field can be solved from the equation of motion, $\tilde \phi  =\Omega^{-1} \phi $
with
\begin{equation}
\label{eq:sol-scalar}
\phi =\frac{c_1 }{2 M}\ln \left(1-\frac{2 M}{r}\right)+c_2.
\end{equation}
where $c_1$ and $c_2$ are integral constants.
It notes that $\phi$ diverges at horizon $r_{\rm H}=2M$,
except $c_1=0$.
Thus to avoid divergency, one could choose constants,
such that $\phi = c_2$.
The corresponding contributions are 
\begin{equation}
I_{\rm GHY}=-\frac{2\pp}{3}   \beta  (2 r-3 M) \phi^2,\qquad
I_{0}=\Omega^{-1} \tilde I_{0}=-\frac{4\pp}{3}  \beta r \phi^2\sqrt{1-\frac{2 M}{r}}, 
\end{equation}
and
\begin{equation}
 \tilde I_{\rm GHY}=\frac{2 \pp \Omega^{2}\beta  \tilde \phi^2}{3 \left(L^2+r^2\right)}
\left[ L^2 (-12 M n+3 M+6 n r-2 r)+r^2 (3 M-2 r)\right].
\end{equation}
Summarizing all above formulas, 
one obtains the Euclidean action 
\begin{equation}
I_{\rm cl}=\tilde I_{\rm cl} =\frac{2\pp}{3}   \beta  c_2^2 M+O\left(r^{-1}\right),
\end{equation}
where in both cases $\beta =8 \pp  M$.
Therefore the internal energy and entropy can be carried out
\begin{equation}
\braket{U}= \frac{4}{3} \pp  c_2^2 M,\qquad
S=\frac{4}{3} \pp ^2 c_2^2 r_{\rm H}^2.
\end{equation}
Restoring the correct factor by $6\equiv 8\pp$ and the field by $\phi^2=c_2^2$,
one can arrive at 
$\braket{U}= \phi^2 M$ and
$S=\phi^2 \pp  r_{\rm H}^2$.
It signifies that the entropy is conformal invariant.
In fact, if one starts with a partition function, 
conformally symmetric at the quantum level, i.e.,
both Lagrangian and measure of functional integral are invariant 
under a conformal change,
all the thermodynamical properties, 
such as temperature and entropy must also be conformal invariant.

\section{The Wald entropy}

Now we check the entropies of above models by Wald's Noether-charge approach.
Since the model Eq.~\eqref{eq:metric} is of simple Lagrangian $R$, 
its entropy density can be read off directly from the Table I in Ref.~\cite{Jacobson:1993vj};
while the conformal RHBs corresponds to the case $F(\phi, R)$, 
and the entropy density is $\partial_R F(\phi, R)=\phi^2$.
The conformal invariance of the entropy can be shown below.
Starting with $2\mathcal{L}=\phi^2 R/(8\pp)-\phi\Box \phi$,
one can find \cite{Brustein:2007jj}
\begin{equation}
\begin{split}
S_{\rm W}&=-\frac{1}{2} \oint_{r=r_{\rm H}, t=\const}
\phi^2 \left(\frac{\delta R}{\delta R_{trtr}}\right)^{(0)}
r^2 \dif \Sigma^2\\
&=-\frac{1}{4} \oint_{r=r_{\rm H}, t=\const}
\phi^2 g^{rr}g^{tt}
r^2 \dif \Sigma^2\\
& = \frac{1}{4} \oint_{r=r_{\rm H}, t=\const}
\phi^2 
r^2 \dif \Sigma^2,
\end{split}\end{equation}
which is conformal invariant by $\phi\to \Omega^{-1} \phi$ and $r^2 \dif \Sigma^2_2 \to \Omega^2 r^2 \dif \Sigma^2_2$.
For $d$-dimensional case,
the scalar field and the area transform 
via $\phi\to \Omega^{(2-d)/2}\phi$ and $r^{d-2}\dif \Sigma^2_{d-2}\to \Omega^{d-2}r^{d-2}\dif \Sigma^2_{d-2}$ respectively. 
Ultimately, we arrive at $S_{W}=\phi^2(r_{\rm H})A_{d-2}r_{\rm H}^{d-2}/4$,
where $\phi(r_{\rm H})$ should be understood as reasonably classical solution of scalar field at horizon.
As to the EGB theory, 
one can obtain from the original action Eq.~\eqref{eq:action-egb} that 
\begin{equation}
\begin{split}
S_{\rm W}&= \frac{A_{d-2}r_{\rm H}^{d-2}}{4}
-\frac{\alpha}{2}\oint_{r=r_{\rm H}, t=\const} 
\left(\frac{\delta B}{\delta R_{rtrt}}\right)^{(0)}
r^{d-2}\dif \Sigma_{d-2}\\
&=\frac{A_{d-2}r_{\rm H}^{d-2}}{4}\left[1-
\frac{2\alpha (d-3) (d-2) (f-1)}{r^2}
\right]_{r=r_{\rm H}}.
\end{split}
\end{equation}
Since $f(r_{\rm H})=0$, 
it will recover the result obtained by path-integral method, Eq.~\eqref{eq:egb-entropy}.

\section{Topology of regular black holes}

Although the conformal RBHs and RBHs with single shape function are all regular from 
the perspective of finiteness of curvature invariants and completeness of geodesics,
they have several different features. 
To distinguish these two classes of RBHs, 
we dub the latter as RBH1, the former as RBH2.

First of all, by absorbing the conformal factor into radius $r\to \Omega^{-1} \rho$,
one can rewrite the metric of RBH2, 
whose product of first two components $g_{tt}g_{\rho\rho}$ is
no longer a constant.
In other words, RBH2 are RBHs with two different shape functions.

Second, the total Ricci curvature and GB scalar of RBH1 is finite, 
i.e., the integrals of $R$ and $B$ over entire manifold are not divergent constants.
As a matter of fact,
given the metric Eq.~\eqref{eq:metric}
one can define two functions
\begin{equation}
\xi(r)=\frac{1}{4M}\int\dif r \int_{0}^{\pp}\dif\theta\sqrt{-g} \; R,
\qquad
\chi(r)=\frac{1}{32  M^2}\int\dif r \int_{0}^{\pp}\dif\theta\sqrt{-g} \; B.
\end{equation}
The both integrations can be taken off, 
\begin{equation}
\xi=r \sigma '+\sigma ,\qquad
\chi=\frac{ \sigma  }{r^3}\left(r \sigma '-\sigma \right).
\end{equation}
Thus the total Ricci curvature and GB scalar are separately
\begin{equation}
\xi_e=\xi(\infty)-\xi(\varnothing),\qquad
\chi_e=\chi(\infty)-\chi(\varnothing),
\end{equation}
where $\varnothing$ denotes the lower boundary.
For $\varnothing=0$, taking into account $\sigma\sim O(r^n)$ with $n\ge 3$ 
as $r\to 0$ and $\sigma \sim O(1)$ as $r\to \infty$,
one has $\xi(0)=\chi(0)=\chi(\infty)=0$ and then 
 concludes that $\xi_e=\sigma(\infty)$ and $\chi_e=0$.
For $\varnothing=r_+$, 
one gets $\xi_e= r_+ \sigma'(r_+)+ 2M/r_+$ 
and $\chi_e=-\left[1-2 M \sigma '\left(r_+\right)\right]/(4 M^2 r_+)$.
While for RBH2, all these topological quantities are divergent, 
because the integration measure is  divergent at center.

Moreover, the different topologies for RHB1 and RHB2
 give a strong suggestion to distinguish these two types of RHBs.
 The pathological topology of RHB2 may shed light on the explanation, 
 why the energy conditions are violated in the entire parameter space \cite{Toshmatov:2017kmw}.
 Therefore finding a conformal RHB without pathological topology and violation of energy conditions
 are still an open problem.

\section{Conclusions and disscusions}

Although the entropy/area law is verified for RHBs by approximated methods or
phenomenological demonstrations, see e.g.\  Refs.~\cite{Myung:2007av,Banerjee:2008gc,Park:2008ud,Kim:2008vi,Dymnikova:2010zz,Miao:2016ulg,Fan:2016hvf},
we provide a solid evidence that entropy/area law 
still holds from path integral and Noether-charge approaches.
This result is against the conditional entropy $S_c$ widely used before,
of which the motivation comes from absorbing the anomaly of mass term into entropy,
rather than the internal energy.
 As to the anomalies in first law (or Smarr formula) with respect to charge or cosmological constant,
 which are regarded as real extension of phase space,
 they can be canceled completely by introducing the pressure of Gliner vacuum at horizon \cite{Lan:2021ngq}.
 The path-integral method also supports this idea, i.e.,
 if the thermal volume of RBHs is applied, 
 the Gliner pressure can be derived from partition function.

 The RHBs are originally created to conquer the UV incompleteness,
 and indeed the classical divergence can be removed,
from the perspective of  finiteness of curvature invariants 
and completeness of geodesics.
From quantum level, it is known that 
the one-loop correction of entropy for singular BHs has logarithm divergent at UV limit \cite{Solodukhin:2011gn}.
Thus it is reasonable to check whether UV divergence still exist in RHBs.
Intuitively, since the essential singularity is removed, 
the coordinate singularity can also be removed by choosing a well-defined coordinates.
As a result, UV divergence may be improved in one-loop correction. 
We will show this in the near future.

 \appendix
 
 \section{The isometric process of Reissner-Nordstr\"om black hole}
 \label{app:rnbh}

 First let us list the some facts for RN black hole, i.e.\
 the entropy, temperature 
 \begin{equation}
 S=\pp r_{\rm H}^2 ,\qquad
 \beta=\frac{2\pp r_{\rm H}^3}{M r_{\rm H}-Q^2},
 \label{eq:tem}
 \end{equation}
and the total action \cite{Gibbons:1976ue}
\begin{equation}
I_{\rm cl}=\frac{\beta}{2}(M-Q \Phi_{\rm H})=\pp r_{\rm H}^2.
\end{equation}
where $\Phi_{\rm H}=Q/r_{\rm H}$.
Then one can work out the {\em enthalphy}
%\begin{equation}
%\braket{U}=M -\frac{Q^2}{r_{\rm H}}
%\end{equation}
\begin{equation}
\mathcal{H}=M -\frac{Q^2}{r_{\rm H}}
\end{equation}
by substituting all above facts into 
\begin{equation}
\label{eq:def-entropy}
S=\beta \mathcal{H}-I_{\rm cl}.
\end{equation}
The Smarr formula can be recovered by
\begin{equation}
\label{eq:smarr}
\frac{I_{\rm cl}}{\beta}=G=\mathcal{H}-\frac{S}{\beta},
\end{equation}
where $G$ is Gibbs energy.
Alternatively, one can treat Eq.\eqref{eq:def-entropy} 
as a differential equation of $\beta(r_{\rm H})$
because of $\mathcal{H}\equiv \partial I_{\rm cl}/\partial \beta$.
The solution of the differential equation is then
\begin{equation}
\label{eq:sol-tem}
\beta= c_1 r_{\rm H}.
\end{equation}
Here $c_1$ is an integration constant.
It implies that when one uses $\mathcal{H}\equiv \partial I_{\rm cl}/\partial \beta$
to calculate the enthalphy, $c_1$ is supposed to be a constant, i.e.
\begin{equation}
\mathcal{H}\equiv \left(\frac{\partial I_{\rm cl}}{\partial \beta}\right)_{c_1}.
\end{equation}
Equating Eq.\eqref{eq:sol-tem} and  Eq.\eqref{eq:tem}, 
one arrives at
\begin{equation}
1-\frac{4\pp}{c_1}=\Phi_{\rm H}^2.
\end{equation}
The $c_1$ be constant signifies that the potential 
$\Phi_{\rm H}$ is fixed when one calculates 
the partial derivative. 
%This is not surprised, since the relation $\dif G=-S \dif T-Q\dif \Phi_{\rm H}$ must hold.

Now we apply the isometric process to recalculate the internal energy and entropy of RN black hole.
Removing the contribution of vector field, 
one has an irreducible action $I_{\rm ir}=M\beta/2$.
To calculate the physical quantities further, 
one can make a replacement
\begin{equation}
Q\to r_{\rm H} \sqrt{1-\frac{4 \pp  r_{\rm H} }{\beta }},\qquad
M\to r_{\rm H}  \left(1 -\frac{2 \pp  r_{\rm H}}{\beta } \right),
\end{equation}
which is obtained similarly to Eq.~\eqref{eq:constraint}.
Thus the classical action becomes the same as Eq.~\eqref{eq:action-hayward},
in turn, one obtains the internal energy $\braket{U}=M_{\rm ir}$ with a fixed thermal volume.
Here $M_{\rm ir}=r_{\rm H}/2$ refers to irreducible mass \cite{Christodoulou:1970wf},
i.e.\ the charge-independent mass,
and connects with mass $M$ via $M=M_{\rm ir}+Q^2/(4M_{\rm ir})$.
It's this quantity that has one-to-one correspondence with surface area \cite{Christodoulou:1972kt}.
As a result, one will recover the entropy/area law in the {\em isometric process} by Eq.~\eqref{eq:def-entropy} and
Smarr formula by Eq.~\eqref{eq:smarr}.
%Alternatively, one has
%\begin{equation}
%\braket{S}_{Q}=\pp  r_{\rm H}^2+\frac{2 \pp  Q^2 r_{\rm H}^2}{r_{\rm H}^2-3 Q^2},\qquad
%\braket{S}_{M}=0
%\end{equation}
%for different processes respectively.

To recover the correct entropy and Smarr formula in the isometric process, 
we have removed the matter sector, 
and calculated all the quantities based on the irreducible sector.
The motivation comes from the RBHs.
For a given regular metric, 
the action of matter sector is not unique, or sometimes is unknown.
However, the physical quantities of one BH with different parameterization, 
such as Eq.\eqref{eq:hayward}, should be the same, 
no matter what kind of matter sector is introduced.
It reflects a fact that the entropy of BH should not depend on the matter sector.

\begin{acknowledgments}
The authors are grateful to Hao Yang (Nankai University) for valuable remarks.
This work was supported in part by the National Natural Science Foundation of China under Grant No.~11675081.
\end{acknowledgments}

\bibliographystyle{apsrev4-2}

\bibliography{references}

\end{document}